# How Creative Ideas Take Shape

Liane Gabora

Department of Psychology, University of British Columbia
Okanagan Campus, Arts Building, 333 University Way, Kelowna BC, V1V 1V7, CANADA

**Abstract**
According to the honing theory of creativity, creative thought works not on individually considered, discrete, predefined representations but on a contextually-elicited amalgam of items which exist in a state of potentiality and may not be readily separable. This leads to the prediction that analogy making proceeds not by mapping correspondences from candidate sources to target, as predicted by the structure mapping theory of analogy, but by weeding out non-correspondences, thereby whittling away at potentiality. Participants were given an analogy problem, interrupted before they had time to solve it, and asked to write down what they had by way of a solution. Naïve judges categorized responses as significantly more supportive of the predictions of honing theory than those of structure mapping.

Creative people often work on a task for a long time before they have insight into how to solve or go about the task, and even if insight occurs it can still take time for the idea can to mature. It is widely assumed that the creative process involves searching through memory and/or selecting amongst a set of predefined candidate ideas. For example, computer scientists have modeled the creative process as heuristic search through a 'state space', a bunch of pre-defined possibilities. Most well known psychological theories of creativity (such as the Geneplore and the Darwinian theory) involve two stages: divergent generation of possibilities, followed by exploration and ultimately selective retention of the most promising of them. In psychological theorizing about creativity there is much discussion of the role of divergent thinking in creativity. Divergent thinking is presumed to involve the generation of multiple, often unconventional possibilities. Thus construed, it necessarily goes hand-in-hand with selection, since if you come up with multiple alternatives you eventually weed some of them out.

However, the generation stage of creative thinking may be divergent not in the sense that it moves in multiple directions or generates multiple possibilities, but in the sense that it produces a raw idea that is vague or unfocused, that requires further processing to become viable. Similarly, the exploration stage of creative thinking may be convergent not in the sense that it entails selecting from amongst alternatives but in the sense that it entails considering a vague idea from different perspectives until it comes into focus. Thus, the terms divergent and convergent may be applicable to creative thought not in the sense of going from one to many or from many to one, but in the sense of going from well-defined to ill-defined, and vice versa. Although a creative process may involve search or selection amongst multiple possibilities, it need not, and neither search nor selection need figure prominently in a general theory of creativity. These are two very different views of creativity; indeed It has been shown mathematically that selection amongst multiple well-defined entities entails a different formal structure from actualizing the potential of a single, ill-defined entity.

There are many reasons to think that creativity involves, in the general case, not selection amongst multiple ideas but the honing of a half-baked idea. A key source of support comes from work on concept combination. A model of concepts that includes potentiality states as a central notion accurately predicts the shifts in applicabilities of properties of concepts when they are placed in different contexts. Studies of concept combination show that the more dissimilar the contributing concepts, the more original, yet potentially the less practical, the resulting idea. This suggests that the more ill-defined the unborn idea, the greater the extent to which it exists in a state of potentiality, and the more processing it requires to become viable. The notion of a state of potentiality comes from the concept of 'ground state' in physics, and the notion of actualization comes from the concept of 'collapse'. Real-time studies of artists and designers indicate that creative ideation involves elaborating on a 'kernel idea', which goes from ill-defined to well-defined through an interaction between artist and artwork. This too is highly consistent with the notion of potentiality states. Indeed it is not obvious how one could simultaneously hold in mind multiple well-defined ideas.

But perhaps the strongest reason is that it is consistent with the structure of associative memory. Because of the ingenious way items are encoded in memory, knowledge and memories that are relevant to the situation or task at hand naturally come to mind (for details see Gabora, 2010). Neural cell assemblies that respond to the particular features of a situation are activated, and items previously encoded in these cell assemblies (that have similar constellations of features and activate similar distributed sets of neurons), are evoked. Both the vagueness of a 'half-baked' idea and the sense that it holds potential, as well as its capacity to actualize in different ways depending on how one thinks it through, may be side effects of the phenomenon of interference. In interference, a recent memory interferes with the capacity to recall an older memory. (A similar phenomenon, sometimes referred to as crosstalk, occurs in neural network models of cognition.) Interference is generally thought of as detrimental, but it may be of help with respect to creativity. A half-baked idea may be what results when two or more items encoded in overlapping distributions of neural cell assemblies interfere with each other and get evoked simultaneously.

The phenomenon of interference leading to creative ideation is referred to as creative interference. When an idea emerges through creative interference, the contributing items are not searched or selected amongst because together they form a single structure. This structure can be said to be in a state of potentiality because its ill-defined elements could take on different values depending on how the analogy unfolds. It is proposed that this unfolding involves disentangling the relevant features from the irrelevant features by observing how the idea looks from sequentially considered perspectives. In other words, one observes how it interacts with various contexts, either internally generated (think it through) or externally generated (try it out).

**A Testbed for the 'Actualization of Potentiality' View of Creativity: Analogy Problem Solving**

To test this new view of creativity as not search or selection but actualization of potential, it was necessary to find a domain in which creativity occurs, and empirically test how people behave in tasks in this domain. Since analogy problem solving is central and ubiquitous facet of human creativity, it seemed like a good choice for a domain.
Analogies are widely believed to involve two elements: a source, which is well understood, and a target, which is less well understood but which shares elements with the source. One might suppose that because analogy does not (in general) require that one come up with something new so much as find a source in memory that is structurally similar to the target, it is the creative process most likely to involve search or selection. Thus if we can show that even analogy problem solving involves not search or selection amongst predefined alternatives but the resolution of ill-defined states of potentiality, we have fairly strong evidence for the hypothesis that potentiality states figure prominently in the creative process.

In tests of analogy solving, the target is presented as a problem that can be solved drawing from the source (Gick & Holyoak, 1983). It is believed that the source tacitly informs the participant, prompting a solution. Let's first look at a preeminent theory of how this happens known as structure mapping. Then we'll look at an alternative that involves the resolution of cognitive states of potentiality.

**Two Views of Analogy**

Structure mapping, in its original formulation, posits that analogy generation occurs in two steps (Gentner, 1983). The first step involves finding an appropriate source in memory and aligning it with the target. The second step involves mapping the correct one-to-one correspondences between the source and the target. Thus structure mapping assumes that once the correct source is found, the analogy making process occurs in isolation from the rest of the contents of the mind. A key principle of structure mapping is the systematicity principle, according to which people prefer to connect structures composed of related predicates.

According to more recent formulation of the theory, the process occurs in three stages. The first stage entails finding all possible source-to-target matches through a quick and dirty process that emphasizes surface similarity. Structural alignment and mapping occur in the second and third stages.

While the basic principles of structure mapping are sensible and well-supported, some of its underlying assumptions have been called into question. First, I propose that the first phase is not, as Gentner (2010) puts it, a "structurally blind free-for-all" (p. 753) but rather, constrained by content-addressable structure of associative memory to naturally retrieve items that are in some way (although not necessarily the right or most relevant way) structurally similar (Gabora, 2000, 2002, 2005, 2010).

A related assumption is that alignment and mapping work with discrete, predefined structures. I proposed that a 'half-baked' analogy exists in a state of potentiality due to creative interference. The source may be an amalgam of multiple items that have previously been encoded to the neural cell assemblies activated by the target, and that in the present context cannot be readily separated from one another. The analogy is in a potentiality state because relevant aspects of these items have not yet been disambiguated from aspects that are irrelevant.

**Contrasting the Predictions of Structure Mapping versus Actualizing Potentiality**

We have looked at two theories concerning the creative process of analogy making. Structure mapping is related to search/selection theories of creativity in emphasizing the challenge of finding a suitable predefined structure. In contrast, according to the 'actualization of potentiality' view (because of the ingenious way that information is encoded in memory) suitable pre-defined structures come to mind for free, not as separate and discrete items but merged together, and the challenge is dis-entangling the relevant from the irrelevant.

The two theories give different predictions as to the state of the mind midway through analogy formation. This is schematically illustrated in Figure 1. Structure mapping predicts that early on, multiple distinct and separate sources may be identified. Eventually an appropriate source is chosen, but the analogy is still unfinished because not all the correspondences between source and target have been found. There is no reason to expect that the incomplete solution will contain extra material.

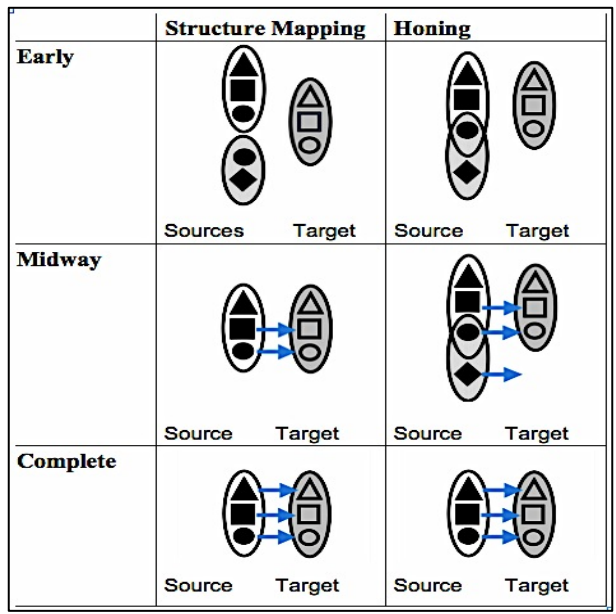

*Figure 1.* Highly simplified illustration of the differences between analogy solving by structure mapping versus honing. According to structure mapping, midway through the analogy making process a correct source has been found but not all the correspondences have been mapped. According to honing, midway through the analogy making process potentially relevant sources have been amalgamated and not all irrelevant correspondences have been weeded out. Note that once the analogy is complete it is not possible to distinguish between the two theories.

According to the 'actualization of potentiality' view, however, the incomplete solution is expected to contain extra material that would perhaps be correct for similar problems but that is not appropriate for this one. The unfinished solution is ill-defined because irrelevant characteristics of the contributing sources have not yet been disambiguated from characteristics that are relevant. The first figure provides a highly simplified illustration of the relevant differences between analogy solving by structure mapping versus by actualization of potentiality.

Note that for completed analogies it is not possible to distinguish which theory provides a better explanation of how the analogy was produced. Only for incomplete analogies do the two theories give different predictions.

**Method**

Adam Saab (a UBC undergraduate student) and I conducted a study to test of the hypothesis that midway through analogy problem solving the mind is in a state of potentiality. (If you're not interested in how the study was conducted, skip ahead to 'Results'.)

Eighty-five University of British Columbia undergraduates who were taking first year psychology course participated in the study. The source and target for this analogy solving experiment were, respectively, The General, and the Radiation Problem, commonly used one-paragraph-long stories in the analogy literature (Gick & Holyoak, 1983). The General involves a fortress that cannot be captured if all soldiers come from the same direction but that can be successfully captured by dividing the army into small groups of soldiers that converge on the fortress from multiple directions. (The complete story is provided in Appendix A). The Radiation Problem involves finding a way to destroy a tumor without killing surrounding tissue. (The complete story is provided in Appendix B). The solution to the Radiation Problem is analogous to the solution to The General; the tumor is destroyed using multiple low-intensity X-rays from different directions.

The experimental procedure consisted of the following: In Phase One, the exposure to source phase, the participants were given five minutes to study The General. They were asked to summarize the story as a test of their story comprehension.

In Phase Two, the problem solving phase, the participants were presented with the target, the Radiation Problem. They were given no indication that the story from phase one could help them solve the problem. Since pilot studies showed that the minimum time required to solve the Radiation problem is two minutes, the participants were interrupted after 100 seconds and told they had 20 seconds to write down whatever they were currently thinking in terms of a solution.

In a follow-up questionnaire distributed immediately afterward, the participants were asked whether they noticed a relation between The General and The Radiation Problem, and if so, at what point they noticed it.

**Judging and Analysis**

Results of both phases were assessed by six judges who were naïve as to the theoretical rationale for the experiment. The story summaries produced in phase one were judged for comprehension on a three-point scale: poor, fair, or good.

Table 1: Elements of The General (source) that are mapped to the Radiation Problem (target) in a successful solution.

| The General (Source) | Radiation Problem (Target) |
|---|---|
| Multiple groups of soldiers | Multiple rays |
| Small groups | Low intensity |
| Groups converge from different directions | Rayss converge from different directions |

Since we were interested in the nature of cognitive states midway through a creative process, participants who correctly solved the problem in the allotted time were removed from the analysis. They were deemed to have correctly solved the problem if they found all three of the correspondences provided in the first table (Table 1).

The judges were asked to categorize each of the remaining incomplete solutions as either Structure Mapping (SM) or Actualization of Potentiality (AP) according to the characteristics of each provided in the second table (Table 2).

Table 2: Characteristics used by naïve judges to categorize artists' responses to questions about their art-making process as indicative of Theory S or Theory H.

| Characteristic | Theory S | Theory H |
|---|---|---|
| If multiple ideas are given, they are | Distinct (e.g., complete ideas separated by 'or') | Jumbled together (e.g., idea fragments spliced together) |
| Ideas are | Well-defined; need to be tweaked and selected amongst | Ill-defined; need to be made concrete; later elements emerge from earlier ones |
| Common core to ideas? | Never | Yes or sometimes |
| Emergent properties? | No | Yes |
| Emergent self-understanding? | No | Yes |
| Emphasis | External product | Internal transformation |

A potential concern at this point is that an answer might contain extraneous information because it was elaborated following retrieval, rather than because of creative interference. There is evidence that analogy making does in some cases involve adapting or elaborating the source to improve the match. Structure mapping does not emphasize this kind of adaptation or elaboration, but it allows for and is not incompatible with it. However, in the analogy used here, no adapting or elaborating of the source was needed to generate a complete and correct solution. In other words, if the correct source (The General story) was found, it could be used as is, without elaboration. Therefore, if extraneous information is present we have good reason to believe that it was due to creative interference.

An example of an answer that was categorized as structure mapping (SM) is:

> "No idea. Don't know much about science. Maybe try to have a low-intensity ray that would sufficiently kill the tumor but not destroy healthy tissues."

In this answer, one of correspondences has been found (correspondence 2: low intensity ray). Since the other two correspondences were not found (multiple rays and different directions) the solution is incomplete. Since the answer provides no evidence that the participant's current conception of a solution consists of multiple items jumbled together in memory, it was classified as SM.

Here are some examples of answers that were categorized as AP. The first example is:

> "First, what kind of tissue will be destroyed with the ray treatment? Can it be replaced using skin graft? How much tissue will be destroyed in the surrounding area? Will the cost outweigh the benefit? This needs to be considered if using the full strength ray."

In this incomplete solution, none of the three correspondences has been found. However, the phrase "if using the full strength ray" indicates that the participant has considered, or is about to consider, the possibility of using a ray that is less than full strength, which suggests that correspondence 2 is within reach. It was classified as AP because it includes irrelevant information (such as concern about the kind of tissue) activated by the target that is unnecessary to generation of the correct solution.

A second example of an answer that was categorized as AP is the following:

> "The high intensity ray is necessary to kill the tumor so maybe shooting it in short successive bursts from different angles will kill the tumor without killing too much healthy tissue."

In this incomplete solution, one of the correspondences has been found: different directions (correspondence 3). It was classified as AP because it also includes irrelevant information (the notion of "short successive bursts") activated by the target that is unnecessary to generation of the correct solution.

**Results**

Each of the 51 incomplete solutions (those that remained after complete solutions were removed from the analysis) was classified as supportive of structure mapping if 4 or more judges judged it as structure mapping, and as supportive of actualizing potentiality if 4 or more judges judged it as actualizing potentiality. In cases where judging was tied, random number generation was used to assign case values. As shown in Figure 2, 39 were classified as supportive of actualizing potentiality, and 12 were classified as supportive of structure mapping. The data supported the hypothesis that analogies are generated by actualizing potentiality.

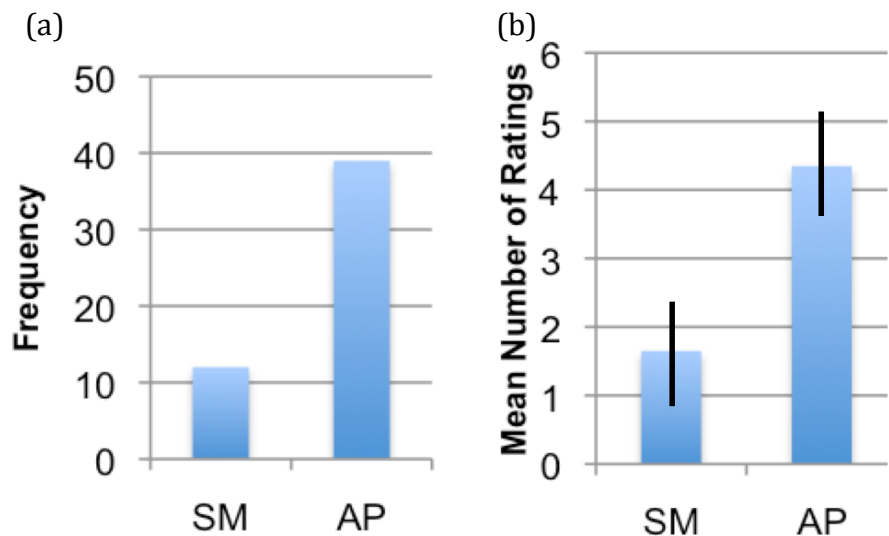

Figure 2. Results of a study in which participants were interrupted midway through an analogy problem-solving task and asked what they were thinking in terms of a solution, and their solutions were classified as indicative of either Structure Mapping (SM) or Honing (HO). (a) Frequency count of solution judgments for SM on the left ($N = 12$) and HO on the right ($N = 39$). (b) Mean number of ratings of SM on the left ($M = 1.65$, $SD = 1.66$) and HO on the right ($M = 4.35$, $SD = 1.66$). Both measures support the prediction of honing theory that midway through creative processing an idea is in a potentiality state (Gabora & Saab, 2011).

A further analysis compared the mean number of judgments (out of a maximum of 6, the total number of judges) across all responses that supported each theory, as shown in Figure 2b. These data corroborate the frequency count findings and provide further support for the AP view of creativity.

**Concluding Remarks**

A few days ago I had the good fortune to speak with Deidre Gentner, the originator of the Structure Mapping theory of analogy, about this project. (By coincidence, we were both at a seminar on problem solving at a castle in Germany that is now used for computer science seminars.) Although framed as an alternative to structure mapping, the data presented here are not incompatible with the key principles of structure mapping, i.e. structural alignment, systematicity, and mapping. We concluded that it would be possible to incorporate the actualization of potential into the structure mapping theory.

In my view, doing so would produce a vastly richer and more accurate model of analogy. The potentiality state is a state in which the entity (in this case the various components of the 'source') can come together in a flexible, context-specific in response to the specific content of the 'target'. The sparse, distributed, content-addressable structure of associative memory ensures that any item that comes to mind as a potential source bears some structural similarity (deep or superficial) to the target. Thus the initial stage of analogy solving is viewed as not "structurally blind", but rather ill-defined due to the multitude of potentially relevant structures that present themselves. Due to the phenomenon of creative interference, what comes to mind may be quite unlike anything that has ever been experienced. Both the data from frequency counts and mean number of SM versus AP judgments support the prediction that a half-baked analogy exists in a state of potentiality, in which the constituent items are merged together, as opposed to a collection of candidate items each in the separate and distinct form in which they were originally encoded in memory. The vagueness or 'half-baked' quality reflects that it is still uncertain how, in the context of each other, these elements come together as a realizable whole.

A limitation of the current study is that it only involved one analogy-solving task. My students and I are investigating the role of potentiality states in other analogy solving tasks, as well as other kinds of creative processes, and in human cognition more generally.

## Acknowledgments

The author is grateful for funding from the Natural Sciences and Engineering Research Council of Canada.

## Appendix A: The General

The following story, referred to as The General, was used as the source in this analogy solving experiment:

> A small country was ruled from a strong fortress by a dictator. The fortress was situated in the middle of the country, surrounded by farms and villages. Many roads led to the fortress through the countryside. A rebel general vowed to capture the fortress. The general knew that an attack by his entire army would capture the fortress. He gathered his army at the head of one of the roads, ready to launch a full-scale direct attack. However, the general then learned that the dictator had planted mines on each of the roads. The mines were set so that small bodies of men could pass over them safely, since the dictator needed to move his troops and workers to and from the fortress. However, any large force would detonate the mines. Not only would this blow up the road, but it would also destroy many neighboring villages. It therefore seemed impossible to capture the fortress. However, the general devised a simple plan. He divided his army into small groups and dispatched each group to the head of a different road. When all was ready he gave the signal and each group marched down a different road. Each group continued down its road to the fortress so that the entire army arrived together at the fortress at the same time. In this way, the general captured the fortress and overthrew the dictator.

## Appendix B: The Radiation Problem

The following story, referred to as The Radiation Problem, was used as the target in this experiment.

> Suppose you are a doctor faced with a patient who has a malignant tumor in his stomach. It is impossible to operate on the patient, but unless the tumor is destroyed the patient will die. There is a kind of ray that can be used to destroy the tumor. If the rays reach the tumor all at once at a sufficiently high intensity, the tumor will be destroyed. Unfortunately, at this intensity the healthy tissue that the rays pass through on the way to the tumor will also be destroyed. At lower intensities the rays are harmless to healthy tissue, but they will not affect the tumor either. What type of procedure might be used to destroy the tumor with the rays, and at the same time avoid destroying the healthy tissue?